\documentclass{article}
\usepackage{times}
\usepackage[left=2.5cm,top=3cm,right=2.5cm,bottom=3cm,bindingoffset=0.5cm]{geometry}
\usepackage{graphicx}
\usepackage[english]{babel}
\usepackage{amsfonts}
\graphicspath{ {illustrations/} }
\usepackage{amsmath}
\usepackage{hyperref}

\begin{document}

% Article top matter
\title{Influence of pulsatile blood flow on allometry of aortic wall shear stress} %\LaTeX is a macro for printing the Latex logo
\author{G. Croizat, A. Kehren, H. Roux de B\'ezieux, A. Barakat\\
	\small{Laboratoire d'hydrodynamique de l'Ecole polytechnique (LADHYX)}\\
	\small{9 Boulevard des Maréchaux, 91120 Palaiseau, France}}
	%\texttt{andyr@comp.leeds.ac.uk}}  %\texttt formats the text to a typewriter style font
\date{}  %\today is replaced with the current date
\maketitle

\begin{abstract}
Shear stress plays an important role in the creation and evolution of atherosclerosis. A key element for in-vivo measurements and extrapolations is the dependence of shear stress on body mass. In the case of a Poiseuille modeling of the blood flow, P. Weinberg and C. Ethier  \cite{wein2006} have shown that shear stress on the aortic endothelium varied like body mass to the power $-\frac{3}{8}$, and was therefore 20-fold higher in mice than in men. However, by considering a more physiological oscillating Poiseuille - Womersley combinated flow in the aorta, we show that results differ notably: at larger masses ($M>10 \ kg$) shear stress varies as body mass to the power $-\frac{1}{8}$ and modifies the man to mouse ratio to 1:8. The allometry and value of temporal gradient of shear stress also change: $\partial\tau/\partial t$ varies as $M^{-3/8}$ instead of $M^{-5/8}$ at larger masses, and the 1:150 ratio from man to mouse becomes 1:61. Lastly, we show that the unsteady component of blood flow does not influence the constant allometry of peak velocity on body mass: $u_{max} \propto M^{0}$. This work extends our knowledge on the dependence of hemodynamic parameters on body mass and paves the way for a more precise extrapolation of in-vivo measurements to humans and bigger mammals.
\end{abstract}

\section*{Introduction}

The formation of atherosclerosis in arteries is a multifactorial process, still not fully understood in its mechanical factors (\cite{mc2011}, ch. 23, p. 502). Yet, it has been shown that wall shear stress plays a key part in this phenomenon \cite{cunn2005}.\\* 
For theoretical analyses as well as for experimentation, it is very useful to know the dependence of wall shear stress $\tau$ in arteries, and especially in the aorta, on body mass $M$. It has long been assumed that wall shear stress was uniform accross species, thus independent of body mass. However, recent works have shown that this assumption was not correct: allometric relationships, expressed as $\tau \sim M^{\alpha}$, with $\alpha \in \mathbb{R}$, can be obtained from simple models of blood flow in the aorta. Note that $\propto$ shall be understood as "is proportional to". P. Weinberg and C. Ethier have modeled blood in the aorta as a Poiseuille flow and have deduced the following result: $\tau_{Poiseuillle} \sim M^{-3/8}$ \cite{wein2006}. This result implies that wall shear stress in mice should be 20 times higher than in men. The objective of this article is to study the allometric relationship between body mass and wall shear stress and other key parameters with a more precise model of blood flow in the aorta. In order to be closer to physiology, we added a purely oscillatory unsteady profile to the steady Poiseuille one. This oscillating component is called Womersley flow.\\*
We will first detail the mathematical derivation of the Poiseuille and Womersley profiles, before exploring their influence on the allometry of wall shear stress (WSS), temporal gradient of wall shear stress (TGWSS) and peak velocity.
\newpage
\section{Mathematical modeling}
In everything that follows, blood is considered as an incompressible Newtonian fluid of density $\rho=1060 \ kg.m^{-3}$ and dynamic viscosity $\mu=3.0 \ 10^{-3} \ Pa.s$. The arteries are assumed to be rigid of radius $a$, and we adopt a no-slip condition on the arterial wall ($u(r=a)=0$). We also assume the flow to be axisymmetric. Model-dependent hypotheses are described in the relevant subsections. See table \ref{tab:scaling} at the end of the document of all constants used in this work.
All numeric simulations were done in Matlab R2016 a, The MathWorks, Inc., Natick, Massachusetts, United States. $log$ stands for logarithm in base 10, and $ln$ for natural logarithm.
\subsection{Poiseuille flow}
In the case of a Poiseuille flow, shear stress on the wall is worth:
\begin{equation}
\tau = 2\mu\frac{u_{max}}{a}
\label{eq:shear_stress}
\end{equation}
Let us detail the allometric arguments of equation \ref{eq:shear_stress}, applied to the aortic artery:
\begin{itemize}
\item{the aortic diameter $a$ is assumed to vary as $M^{0.375}$, both from theoretical (\cite{west1997}) and experimental considerations (\cite{clark1927} and \cite{holt1981})}
\item{the dynamic viscosity of blood $\mu$ is assumed to be independent of the body mass}
\item{the aortic velocity $u_{max}$ can be seen as the cardiac flow rate divided the aortic cross-section $u_{max}=\frac{Q}{\pi a^{2}}$. The cardiac flow rate varies with $M^{0.75}$, as described in \cite{west1997} and \cite{kleib1932}. Therefore, \textbf{$u_{max}$ is independent of $M$.}}  
\end{itemize}
We can conclude that in the case of a Poiseuille flow, wall shear stress scales as:
\begin{equation}
\tau_{Poiseuille} \propto M^{-0.375}
\end{equation}
Additionally, we can deduce the allometry of temporal gradient of wall shear stress. Since $\frac{\partial \tau}{\partial t} \sim \omega\tau$, $\omega \sim M^{-0.25}$ being the cardiac frequency, we get:
\begin{equation}
\frac{\partial \tau}{\partial t}_{Poiseuille} \sim M_b^{-0.625}
\end{equation}
We obtain the following values for mice, rabbits and humans (tab. \ref{tab:cha}), as described in \cite{wein2006}, relatively to the human value: 
\begin{table}[!h]
\centering
\begin{tabular}{|c|c|c|c|c|}
	\hline \textbf{Animal} & \textbf{Mass} & \textbf{Shear stress} & \textbf{Gradient of shear stress} & \textbf{Peak velocity} \\
	\hline Human & 75 kg & 1 & 1 & 1 \\
    \hline Rabbit & 3.5 kg & 3.0 & 6.8 & 1 \\
    \hline Mouse & 25 g & 20.2 & 148.8 & 1 \\
	\hline 
\end{tabular}
\caption{Relative flow characteristics for various species - Poiseuille flow}
\label{tab:cha}
\end{table}
\subsection{Womersley flow}
We consider a purely oscillating pressure gradient (eq. \ref{eq:pres_grad_1}) in the same geometry, and derive the allometry of wall shear stress, temporal gradient of wall shear stress and peak velocity associated to this flow.
\subsubsection{Ruling equations}
In this context, the Navier-Stokes equation projected on $z$ can be written as:
\begin{equation}
\rho\frac{\partial u}{\partial t}=-\nabla P +\mu \Delta u
\label{eq:ns_wom}
\end{equation}
We inject the expression of the pressure gradient (with $G_{1} \in \mathbb{R}$ constant) in \ref{eq:ns_wom}:
\begin{equation}
-\nabla P = \Re(G_{1}\exp(i\omega t))
\label{eq:pres_grad_1}
\end{equation}
as well as the expression of the velocity (note that $A(r) \in \mathbb{C}$):
\begin{equation}
u(r,t) = \Re(A(r) \exp(i\omega t))
\label{eq:wom_init}
\end{equation}
We obtain the following equation, $'$ being the differentiation with respect to $r$.
\begin{equation}
A^{''}+ \frac{1}{r}A^{'}- \frac{i \omega \rho}{\mu} A = -\frac{G_{1}}{\mu}
\label{eq:Bessel}
\end{equation}
where we recognize the sum of a constant particular solution and a Bessel differential equation. We get the following velocity field, with $J_{0}$ the 0-order Bessel function of first kind. We also introduce the Womersley number $\alpha = a\sqrt{\frac{\omega\rho}{\mu}}$. This dimensionless number corresponds to the ratio of inertial transient forces to viscous forces.
\begin{equation}
u(r,t) =\Re(\frac{G_{1}}{i\omega\rho}(1-\frac{J_{0}(i^{\frac{3}{2}}\alpha\frac{r}{a})}{J_{0}(i^{\frac{3}{2}}\alpha)})e^{i\omega t}) \ \ on \ \vec{u_{z}}
\label{eq:wom_veloc}
\end{equation}
\subsubsection{Derivation}
Let us now calculate the flow rate $Q$:
\begin{equation}
Q(t)=\int_{0}^{a}u(r)2\pi rdr
\label{eq:wom_flow}
\end{equation}
and the wall shear stress $\tau$:
\begin{equation}
\tau(t)=-\mu\frac{\partial u}{\partial r}|_{r=a}
\label{eq:wom_shear}
\end{equation}
The calculation of (\ref{eq:wom_shear}) goes down to
\begin{equation}
\frac{\partial}{\partial r}[J_{0}(i^{\frac{3}{2}}\alpha\frac{r}{a})]_{|r=a}
\end{equation}
Using $J_{0}^{'}=J_{1}$ in $\mathbb{C}$, we can conclude that
\begin{equation}
\tau(t)=\Re(\frac{a G_{1}}{i^{\frac{3}{2}}\alpha}\frac{J_{1}(\alpha i^{\frac{3}{2}})}{J_{0}(\alpha i^{\frac{3}{2}})}e^{i\omega t})
\label{eq:wom_shear_2}
\end{equation}
Concerning (\ref{eq:wom_flow}), the key point is
\begin{equation}
\int_{0}^{a}rJ_{0}(i^{\frac{3}{2}}\alpha\frac{r}{a})dr
\end{equation}
Using $x^{n}J_{n-1}(x)= \frac{d}{dx}(x^{n}J_{n}(x))$ for n in $\mathbb{N}$* and x in $\mathbb{C}$, we get
\begin{equation}
Q(t)=\Re(\frac{\pi a^{2}i G_{1}}{\omega \rho}(1-\frac{2J_{1}(\alpha i^{\frac{3}{2}})}{\alpha i^{\frac{3}{2}}J_{0}(\alpha i^{\frac{3}{2}})})e^{i\omega t})
\label{eq:wom_flow_2}
\end{equation}
\subsection{Physiological pulse wave: Poiseuille + Womersley flow}
In reality, the pressure gradient is neither steady, nor purely oscillatory. It was shown (\cite{mor20013}, \cite{wom1955}) that the pressure gradient $\nabla P$ could be reasonably approximated by the first six harmonics of its Fourier decomposition. In our case, six harmonics are not needed, as we are only looking at allometric variations. Along with the allomerty of the Womersley flow (first harmonic), we will present the sum of Poiseuille and Womersley flows (fundamental + first harmonic), \textbf{which we denote PW flow from here on}:
\begin{equation}
u_{PW}=u_{Poiseuille}+u_{Womersley}
\end{equation}
Consequently, by linearity of differentiation, wall shear stress and its temporal gradient are obtained by: $\tau_{PW}=\tau_{Poiseuille}+\tau_{Womersley}$ and $\nabla\tau_{PW}=\nabla\tau_{Poiseuille}+\nabla\tau_{Womersley}$. Concerning peak velocity however, calculations are done using the total velocity field $u_{PW}$.  
% The question is: in what proportion do the Poiseuille and Womersley components appear in the physiological pulse wave of different mammals ? We use the measurements done by Womersley to scale the relative components at human mass: writing the Fourier decomposition of the  pressure gradient:
% \begin{equation}
% \nabla P= \sum_{i=0}^{\infty} G_i \exp(i \omega t)
% \end{equation}
% We recall the respective velocity profiles, expressed as a function of the $G_i$:
% \begin{equation}
% u_{Womersley} =\Re(\frac{G_{1}}{i\omega\rho}(1-\frac{J_{0}(i^{\frac{3}{2}}\alpha\frac{r}{a})}{J_{0}(i^{\frac{3}{2}}\alpha)})e^{i\omega t})
% \end{equation}
% and
% \begin{equation}
% u_{Poiseuille}=\frac{G_0 a^2}{4\mu}(1-(r/a)^2)
% \end{equation}
% In the human aorta, Womersley has measured $G_0 = 1999 \ Pa/m$ and $G_1 = 1534 \ Pa/m$ \cite{wom1955}. Consequently, we obtain $\frac{G_1/i\omega\rho}{G_0 a^2/4\mu}=0.43$. This ratio allows us to scale the two components at the human mass $M=70 \ kg$.
%
\section{Results}
\subsection{Wall shear stress (WSS)}
\begin{figure} [!h]
	\centering
	\includegraphics[width=0.6\textwidth]{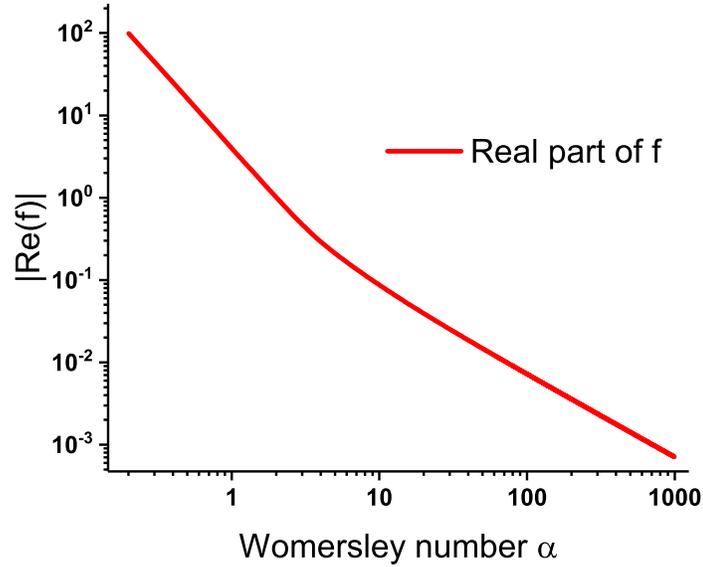}
	\caption{Graph of \large{$log|\Re(f)|$} as a function of $\alpha$, logarithmic scale. We can clearly identify two power laws: $\Re(f) \propto \alpha^{-2}$ for $\alpha < 2$, and $\Re(f) \propto \alpha^{-1}$ for $\alpha > 5$}
	\label{shear_fit}
\end{figure}

\begin{figure} [!h]
	\centering
	\includegraphics[width=0.6\textwidth]{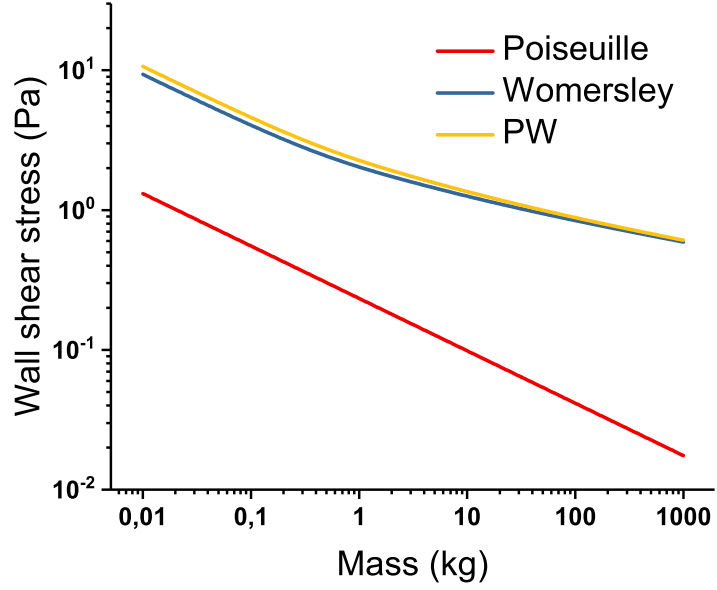}
	\caption{Compared wall shear stresses for Poiseuille, Womersley and PW flows as a function of body mass - logarithmic scale. We observe that Poiseuille stress is globally dominated, in absolute value, by Womersley stress: Poiseuille makes only $12\%$ of total stress for a mouse ($M=0.025 \ kg$), and this percentage goes down to $5\%$ for a human ($M=75 \ kg$). Concerning allometry, both processes induce $\tau_{PW} \propto M^{-0.375}$ at low masses ($M < 1kg$), but at higher masses ($M>10kg$), the Womersley stress sets the allometry $\tau_{PW} \propto M^{-0.125}$.}% Equality of stresses assumed for $\alpha = 10 \ \leftrightarrow M=27 \ kg$.}
	\label{fig: compared_wss}
\end{figure}

To evaluate the allometry of wall shear stress, we need to know the allometry of every variable present in its expression. Concerning $a$, $\alpha$ or $\omega$, the allometry is known from \cite{wein2006}, as summed up in table \ref{tab:scaling}. But the allometry of $G_{1}$ is unknown. To overcome this problem, we substitute the flow rate $Q$, which follows an experimentally determined allometry $Q \sim M^{0.75}$. This step can be discussed, since the Womersley flow model implies a zero average flow rate. Although $<Q>$, the time average value of Q, is always null, its effective value $<Q^2>$ is indeed an increasing function of body mass, which justifies the use of $Q \propto M^{0.75}$ in the Womersley model.
% In section 3 (spectral decomposition of the pressure gradient), it is no longer be possible to substitute the flow rate, and an alternative approach is described. Note that this substitution also absorbs the time dependence of the velocity. 
Combining (\ref{eq:wom_shear_2}) and (\ref{eq:wom_flow_2}), we obtain
\begin{equation}
\underline{\tau}=\frac{\omega\rho J_{1}(x)}{\pi a i(xJ_{0}(x)-2J_{1}(x))}\underline{Q}
\label{eq:tau_wom}
\end{equation}
with $x=i^\frac{3}{2}\alpha$ and $\underline{Q}$, $\underline{\tau}$ the complex variables such that $Q=\Re(\underline{Q})$ and $\tau=\Re(\underline{\tau})$.\\
We introduce the following function $f$, such that $\underline{\tau} = \frac{\omega \rho}{\pi a} f(\alpha) \underline{Q}$:
\begin{equation}
f(\alpha)= \frac{iJ_{1}(x)}{(2J_{1}(x)-xJ_{0}(x))}
\end{equation}
We plot $|\Re(f)|$ as a function of $\alpha$  in logarithmic scale on figure \ref{shear_fit}). Note that $\Re(f)$ is negative, ie $|\Re(f)|=-\Re(f)$.
\begin{itemize}
\item for $\alpha < 2$, $\Re(f) \sim \alpha^{-2}$ ($r^{2} = 0.98$)
\item for $\alpha > 5$, $\Re(f) \sim \alpha^{-1}$ ($r^{2} = 0.99$)
\end{itemize}
We can confirm these regressions using Taylor and asymptotic developments of $J_0$ and $J_1$:
\begin{equation}
J_{0}(x)=1-x^2/4+O_{0}(x^4)
\label{eq:J0_low}
\end{equation}
and
\begin{equation}
J_{1}(x)=x/2-x^3/16+O_{0}(x^5)
\label{eq:J1_low}
\end{equation}
using big O notation (\cite{arf2015}, p. 687).\\
Consequently, for $\alpha <<1$, $f(\alpha) \ \approx \ 4i/x^2 \approx -4/\alpha^2$, which confirms $\Re(f) \sim \alpha^{-2}$ for $\alpha << 1$.\\
Conversely, for $|x|\rightarrow +\infty$:
\begin{equation}
J_{0}(x)\approx \sqrt{\frac{2}{\pi x}}cos(x-\pi/4)
\label{eq:J0_high}
\end{equation}
and
\begin{equation}
J_{1}(x)\approx \sqrt{\frac{2}{\pi x}}sin(x-\pi/4)
\label{eq:J1_high}
\end{equation}
introducing complex sine and cosine (\cite{arf2015}, p. 723).\\
In our case, $x=i^{3/2}\alpha=\alpha \exp(3i\pi/4)$, therefore $f(\alpha) \approx -\frac
{itan(x)}{x}=\frac{tanh(\alpha\sqrt{2}/2)}{\alpha}\exp(-3i\pi/4)$ and finally $\Re(f(\alpha)) \approx \frac{\sqrt{2}}{2\alpha}$, which confirms $\Re(f) \sim \alpha^{-1}$ for $\alpha \rightarrow \infty$.\\
Knowing that $\alpha \sim M^{0.25}$, and that $\frac{\omega \rho Q}{\pi a} \sim M^{0.125}$ (see table \ref{tab:scaling} at the end of the document), we obtain:\\
\begin{center}
\fbox{$\tau_{womersley} \sim M^{-0.375}$ for $M <1 \ kg$} and \fbox{$\tau_{womersley} \sim M^{-0.125}$ for $M > 10 \ kg$}
\end{center}
which differs notably from
\begin{center}
\fbox{$\tau_{poiseuille} \sim M^{-0.375}$ for all $M$}
\end{center}
We can complete the previous chart of relative shear stress for different animals (tab. \ref{tab:cha_P_vs_Wom}).\\
\begin{table}[!h]
	\centering
	\begin{tabular}{|c|c|c|c|c|c|}
		\hline \textbf{Animal} & \textbf{Mass} & $\alpha_{aorta}$ & \textbf{Shear stress (Poiseuille)} & \textbf{Shear stress (Womersley)} & \textbf{Shear stress (PW)} \\ 
        \hline Human & 75 kg & 13 & 1 & 1 & 1\\
        \hline Rabbit & 3.5 kg & 6 & 3.0 & 1.7 & 1.8\\
		\hline Mouse & 25 g & 1.8 & 20.2 & 7.6 & 8.2\\ 
		\hline 
	\end{tabular}
	\caption{Relative shear stress for various species - Poiseuille, Womersley, and Poiseuille + Womersley (PW) flows}
	\label{tab:cha_P_vs_Wom}
\end{table}
We also plot the compared shear stress induced by a Poiseuille, Womersley or physiological PW flows as a function of mass on figure \ref{fig: compared_wss}. 
First of all, we can see that despite several allometric approximations (flow rate, heart rate, aortic radius, Womersley number), the absolute value of wall shear stress $\tau \in [0.1;10] \ Pa$ is within the range of measured values (\cite{mc2011}, ch. , p.502). We observe that Poiseuille shear stress is, in absolute value, much lower than Womersley stress (10 to 20 times lower as mass increases). At low masses ($M<1 \ kg$), both flows follow the same allometry: $\tau_{PW} \propto M^{-0.375}$. At higher masses however, ($M>10 \ kg$), the Womersley stress accounts for the variations of  total stress and thus $\tau_{PW} \propto M^{-0.125}$. In practice, it means that \textbf{the difference in wall shear stress between mice and men is greatly reduced by the influence of pulsatile flow in larger arteries}, and shows the necessity of the unsteady modeling of blood flow for correct body mass allometries.
\subsection{Temporal gradient of wall shear stress (TGWSS)}
\begin{figure} [!h]
	\centering
	\includegraphics[width=0.6\textwidth]{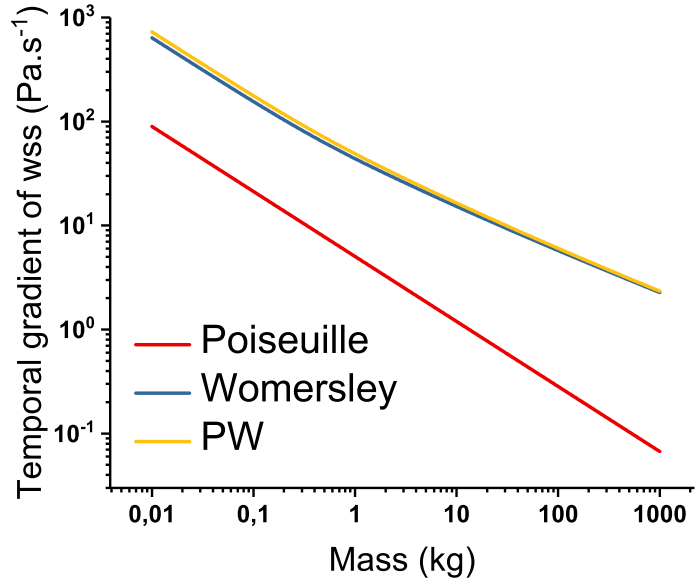}
	\caption{Compared temporal gradient of wall shear stresses for Poiseuille, Womersley and PW flows as a function of body mass - logarithmic scale. Poiseuille TGWSS is much smaller than Womersley TGWSS. At low masses ($M < 1kg$), $\frac{\partial \tau}{\partial t}_{PW} \propto M^{-0.625}$, while at higher masses ($M>10kg$), the Womersley stress accounts for the body mass allometry $\frac{\partial \tau}{\partial t}_{PW} \propto M^{-0.375}$.}% Equality of stresses assumed for $\alpha = 10 \ \leftrightarrow M=27 \ kg$.}
	\label{fig: compared_tpwss}
\end{figure}

Concerning the temporal gradient of shear stress, we get $\frac{\partial \tau}{\partial t}\sim \omega\tau$, with $\omega \sim M^{-0.25}$ \cite{wein2006}, and therefore:\\ 
\begin{center}
\fbox{$\frac{\partial \tau}{\partial t}_{wom} \sim M_b^{-0.625}$ for $M < 1 \ kg$} and \fbox{$\frac{\partial \tau}{\partial t}_{wom} \sim M_b^{-0.375}$ for $M > 10 \ kg$}
\end{center}
Which should be compared to 
\begin{center}
\fbox{$\frac{\partial \tau}{\partial t}_{poi} \sim M^{-0.625}$ for all $M$}
\end{center}
We compare the values of shear stress gradient between Womersley and Poiseuille in tab. \ref{temp_grad}. Poiseuille TGWSS is lower than Womersley TGWSS, and the allometric domination at high masses is also present. The graph of these functions (fig. \ref{fig: compared_tpwss}) is very similar to the one on wall shear stress (fig. \ref{fig: compared_wss}). At low masses ($M< 1kg$), both flows result in $\frac{\partial \tau}{\partial t}_{PW} \propto M^{-0.625}$ allometry, and at higher masses ($M>10kg$), Womersley stresses dominate the allometry: $\frac{\partial \tau}{\partial t}_{PW} \propto M^{-0.375}$. Here again, we observe a reduced difference between men and mice due to the influence of unsteady flow at high Womersley numbers.%
\begin{table}[!h]
	\centering
	\begin{tabular}{|c|c|c|c|}
		\hline \textbf{Animal} & \textbf{Shear stress gradient (Poiseuille)} & \textbf{Shear stress gradient (Womersley)} & \textbf{Shear stress gradient (PW)} \\ 
		\hline Human & 1 & 1 & 1\\ 
		\hline Rabbit & 6.8 & 3.8 & 3.9\\ 
		\hline Mouse & 148.8 & 55.8 & 60.5\\ 
		\hline 
	\end{tabular}
	\caption{Relative shear stress for various species - Poiseuille vs Womersley flow}
	\label{temp_grad}
\end{table}
\subsection{Peak velocity}
\begin{figure}[!h]
\centering
  \includegraphics[width=.75\textwidth]{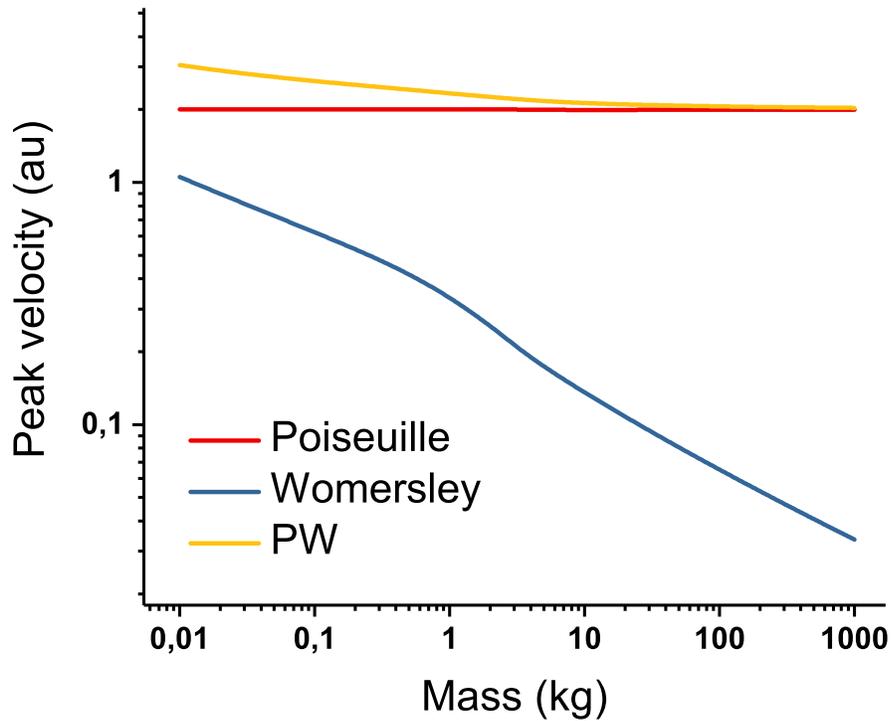}
\caption{Peak velocity of Poiseuille, Womersley and PW flows as a function of mass, logarithmic sclae: the constant peak velocity observed experimentally (\cite{sch1984}) is due to the Poiseuille component rather than the Womersley one which vanishes at high M}
\label{fig:peak_veloc}
\end{figure}

And finally, we can express the maximum velocity. First we inject the expression of the flow rate in the Womersley velocity to absorb the time dependence (again using $x=i^{3/2}\alpha$) :
\begin{equation}
u_{Wom}(r)=\frac{Q}{\pi a^2} \times \frac{J_0 (x r/a)-J_0(x)}{x J_0(x)-2J_1(x)}
\end{equation}
We know that $\frac{Q}{\pi a^2}$ is independent of $\alpha$ (table \ref{tab:scaling}), therefore
\begin{equation}
u^{Wom}_{max}(\alpha) \propto \max_{r\in [0,a]}\Re \Bigg( \frac{J_0 (x r/a)-J_0(x)}{x J_0(x)-2J_1(x)}\Bigg) =\max_{r\in [0,a]}(g_{wom}(\alpha))
\end{equation}
Recalling equivalents of $J_0$ and $J_1$ (eq. \ref{eq:J0_low} to \ref{eq:J1_high}), we can find equivalents of $g_{wom}(\alpha)$ at low and high $\alpha$:
\begin{itemize}
\item for $M << 1$, $g_{wom} \ \sim \frac{\sqrt{2}}{2\alpha} \sim M^{-0.25}$, we observe a divergence: $$\lim_{M\to 0} u^{Wom}_{max} = + \infty$$.
\item for $M >> 1$, we also obtain $g_{wom} \ \sim \frac{1}{\alpha} \sim M^{-0.25}$, and $$\lim_{M\to +\infty} u^{Wom}_{max} = 0$$.
\end{itemize}
We plot $\max_{r}(g_{wom})$, which represents the Womersley peak velocity only, as a function of M on figure \ref{fig:peak_veloc}.  We also add the Poiseuille peak velocity.\\
Concerning PW flow, calculations must be done considering the total velocity field: recalling the expression of $u_{Pois}$ as a function of $Q$ and $r$: $u_{Pois}=\frac{2Q}{\pi a^2}(1-\frac{r}{a})$, $u^{PW}_{max}$ is: 
\begin{equation}
u^{PW}_{max}(\alpha) \propto  \max_{r\in [0,a]} \Re \Bigg( 2(1-\frac{r}{a})+\frac{J_0 (x r/a)-J_0(x)}{x J_0(x)-2J_1(x)} \Bigg) =\max_{r\in [0,a]}(g_{PW}(\alpha))
\end{equation}

On figure \ref{fig:peak_veloc}, we can see that the Womersley flow alone gives rise to a peak velocity which vanishes with $M$ (blue curve), while the PW flow (yellow curve) shows a constant peak velocity for reasonably high $M$. This proves that the constant peak velocity measured experimentally \cite{sch1984} is due to the steady Poiseuille component of the flow, rather than to the Womersley unsteady component. We also observe that the maximum velocity is systematically reached in the center of the artery ($r=0$, data not shown), which is not the case for a pure Womersley flow at high Womersley number. This confirms the domination of the Poiseuille flow. Figure \ref{fig:peak_veloc} further proves the necessity of a combined Poiseuille - Womersley flow to account for the variations of the different hemodynamic parameters.
\subsection{Summary of the different allometric laws}
We summarize the different allometries obtained throughout this work in table \ref{tab:allometries}.\\
\begin{table}[!h]
	\centering
	\begin{tabular}{|l|c|c|c|}
      \hline
      Flow modeling & Poiseuille & Womersley & PW\\
      \hline
       Hemodynamic parameter &  & \begin{tabular}{@{}c|c@{}}$ M<1 \ \ kg$ & $M>10 \ kg$\end{tabular} & \begin{tabular}{@{}c|c@{}} $M<1 \ \ kg$ & $M>10 \ kg$\end{tabular}\\ \hline
      WSS & -0.375 & \begin{tabular}{@{}c|c@{}}$-0.375$ & $-0.125$\end{tabular} & \begin{tabular}{@{}c|c@{}}$-0.375$ & $-0.125$\end{tabular}\\ \hline
      TGWSS & -0.625 & \begin{tabular}{@{}c|c@{}}$-0.625$ & $-0.375$\end{tabular} & \begin{tabular}{@{}c|c@{}}$-0.625$ & $-0.375$\end{tabular}\\ \hline
      Peak velocity & 0 & -0.25 & 0 \\ \hline
      \end{tabular}
	\caption{Summary of body mass allometry of different hemodynamic parameters}
	\label{tab:allometries}
\end{table}
\section*{Conclusion}
%
%
% We have shown that the allometric relationships of wall shear stress and peak velocity with body mass change notably when modeling blood flow as a - more physiologic - Womersley flow rather than a steady Poiseuille flow. The results obtained are closer to experimental evidences for shear stress, but not as precise concerning maximal velocity.\\*
% An interesting future work could me to model the flow more precisely, using a Fourier decomposition in several harmonics, and thus try to find a model that would fit the experimental data on all levels.
In this study, we have recalled the derivation of wall shear stress, temporal gradient of wall shear stress and peak velocity, for a Poiseuille and subsequently Womersley modeling of the blood flow in the aorta.
We have shown that a Poiseuille modeling alone (and equivalently, a Womersley modeling alone) was not able to account for the variations of the previously mentionned parameters. We showed that a combination of Poiseuille and Womersley flows, here called PW flow, was necessary to conduct an allometry study.
\begin{itemize}
\item Concerning wall shear stress, Poiseuille and Womersley flows agree on a $M^{-0.375}$ at low Womersley numbers, while the Womersley component dominates at higher $\alpha$ ($M^{-0.125}$ against $M^{-0.375}$). This changes the "Twenty-fold difference in hemodynamic wall shear stress between murine and human aortas" advanced in \cite{wein2006} to a lower 8.2 fold difference.
\item About the temporal gradient of wall shear stress, the same phenomenon occurs: while $M^{-0.625}$ is valid at low $\alpha$, the unsteady component dominates at higher $\alpha$ ($M^{-0.375}$ against $M^{-0.625}$), and brings mice to men ratio from $150:1$ to $61:1$.
\item For peak velocity on the contrary, the independance on $\alpha$ observed experimentally \cite{sch1984} is due to the Poiseuille component rather than to the Womersley one, which follows a vanishing $u_{max}^{wom} \propto M^{-0.25}$.
\end{itemize}

As a consequence, only considering a Poiseuille flow to infer the wall shear stress allometry in mammals is not a precise method. On the contrary, one should take into account the two components of the blood flow profile (steady Poiseuille and unsteady Womersley), to account for their respective ranges of domination. 
% Some measurements were done on small animals (\cite{gre2006}), which concluded as expected on a Poiseuille-like $\tau \propto M^{-0.38}$ power law, but to our knowledge, no experiments have been conducted on heavier animals, to confirm the influence of the Womersley stress. A comprehensive imaging study of aortic velocity profiles on heavy mammals could greatly benefit the field.
This work gives a new insight on the allometry of 3 key hemodynamic parameters in atherosclerosis. It can also help extrapolate measurements from small mammals like mice, which are common in cardiovascular laboratories, to bigger ones and to humans. A comprehensive imaging study of aortic velocity profiles on heavy mammals could help support or reject this theory and greatly benefit the field.
\section*{Appendix}
\renewcommand{\arraystretch}{1.2}
\begin{table}[!h]
	\begin{center}
		\begin{tabular}{|c|c|c|c|}
		\hline \textbf{Physical quantity} & Body mass allometry & Human value ($M=75 \ kg$) & Scaling constant (IS unit)\\
        \hline Aortic radius & $a_{0}M^{0.375}$ & $1.5 \ cm$ & $a_{0}=3.0 \ 10^{-3} \ m$\\ 
        \hline Flow rate & $Q_{0}M^{0.75}$ & $ 5L/min=8.3 \ 10^{-5}m^3/s $ & $Q_{0}=3.3 \ 10^{-6} \ m^3.s^{-1}$\\ 
        \hline Cardiac pulsation ($2\pi*$ frequency) & $w_{0}M^{-0.25}$ & $7.35 \  rad.s^{-1} \ =70 \ bpm$ & $w_{0}=21.6 \ rad.s^{-1}$\\ 
		\hline Womersley number & $\alpha_{0}M^{0.25}$ & $13$  & $\alpha_{0}=4.4$\\ 
		\hline Blood dynamic viscosity & $\mu M^{0}$ & / & $\mu=3.0 \ 10^{-3} \ Pa.s$\\ 
        \hline Blood density & $\rho M^{0}$ & / & $\rho=1.06 \ 10^{3} \ kg.m^{-3}$\\ 
        \hline 
	\end{tabular}
	\caption{Body mass allometry of different body parameters. The scaling constants were determined to match the value of the human proximal aorta}
	\label{tab:scaling}
    \end{center}
\end{table}
%
 %Must end the environment
%
\end{document}